\documentclass[aps,amsmath,showpacs,amsfonts,11pt]{revtex4}
\usepackage{graphicx}
\usepackage{epstopdf}
\usepackage{epsfig,graphicx}
\usepackage[english]{babel}
\usepackage{amsfonts}
\usepackage{amsmath}
\usepackage{latexsym}
\usepackage{graphics,bm}
\usepackage{dcolumn}
\usepackage{bm}
\usepackage{rotating}

\begin{document}

\title{Heisenberg Operator Approach for Spin Squeezing Dynamics}

\author{Aranya B. Bhattacherjee$^{1}$,  Deepti Sharma$^{1}$, and  Axel Pelster$^{2}$}

\address{$^{1}$ School of Physical Sciences, Jawaharlal Nehru University, New Delhi-110067, India \\
$^{2}$ Fachbereich Physik und Forschungszentrum OPTIMAS, Technische Universit\"{a}t Kaiserslautern, 67663, Germany}

\begin{abstract}
We reconsider the one-axis twisting Hamiltonian, which is commonly used for generating spin squeezing, and treat its dynamics within the Heisenberg operator approach. To this end we solve the underlying Heisenberg equations of motion perturbatively and evaluate the expectation values of the resulting time-dependent Heisenberg operators in order to  determine approximately the dynamics of spin squeezing. Comparing our results with those originating from exact numerics reveals that they are more accurate than the commonly used frozen spin approximation.
\end{abstract}


\maketitle

\section{Introduction}
Atomic spin squeezing is a quantum effect of collective spin systems  \citep{gross,ma} with the potential to improve the precision of measurements in experiments in general \citep{wine1,wine2,polzik,cronin} and to study particle correlations as well as entanglement in particular \citep{sorensen,bigelow,gue}. The quantum mechanical uncertainty of spin operators limits the measurement accuracy of spectroscopic investigations and the performance of atomic fountain clocks \citep{wine1,wine2}. The standard uncertainty relation of angular momentum operators predicts a spectroscopic sensitivity proportional to $1/\sqrt{N}$, where $N$ denotes the total number of atoms utilized in the given spectroscopic investigation. It was suggested in Ref. \citep{kita1} to produce spin-squeezed states, which redistribute the uncertainty unevenly between two components of the total angular momentum, so that measurements, which are sensitive to the component with reduced uncertainty, become more precise. These states were applied in atomic clocks for reducing quantum noise \citep{wine1,wine2,tur,meyer,leib} and for implementing quantum information processing \citep{sorensen,wang,kor,yi}.  Spin-squeezed states can also be experimentally realized in a BEC \citep{orz,ried,est,mue}, which allows for instance to detect weak forces \citep{Jaa}.  In such systems, spin fluctuations in one spin component perpendicular to the mean spin direction turn out to be reduced below the standard quantum limit (SQL).
Quantum correlations among individual spins are responsible for spin squeezing provided that a nonlinear interaction between the spins is present. In the original proposal by Kitagawa and Ueda \citep{kita1}, two fundamental types of nonlinear spin interactions were identified, which are called one-axis twisting and two-axis counter twisting. One-axis twisting interaction is referred to a nonlinear term of the form $\hat{J}_{\alpha}^{2}$, $\alpha=x,y,z$ in the Hamiltonian, while if the twisting is performed
simultaneously clockwise and counterclockwise about two orthogonal axes in the plane normal to the mean spin direction, it is referred to as two-axis counter twisting. The Hamiltonian in the second case then contains a term of the form $\hat{J}_{\alpha} \hat{J}_{\beta}+\hat{J}_{\beta} \hat{J}_{\alpha}$ with $\alpha \neq \beta$. Later on Law et al. \citep{law1} examined spin squeezing in a collection of interacting spins in the presence of an external field, which demonstrated a strong reduction of spin fluctuations that could be maintained for a much longer period of time.
So far, spin squeezing was theoretically studied either exactly by using numerical simulations \citep{jin1,jin2,jin3} or analytically within the frozen spin approximation \citep{law1,bhatt1}. The advantage of the frozen spin approximation relies in the fact that it can straight-forwardly be applied to study many different systems which includes spin dynamics and entanglement in mixed Hamiltonian model \citep{fsa1,fsa2}, Lipkin-Meshkov-Glick model \citep{fsa3,fsa4}, generalized two-axis twisting model \citep{fsa5}, and dipolar condensates \citep{fsa6}.

In this paper, we start with reconsidering the one-axis twisting model of Law et al. \citep{law1} in Section II. To this end we analytically solve the underlying Heisenberg equations of motion for the atomic degrees of freedom by using a perturbative technique and by treating its dynamics within the Heisenberg operator approach. By evaluating the expectation values of the respective operators in the Heisenberg picture, the resulting dynamics of spin squeezing is then determined in Section III. Finally, comparing our analytical results with exact numerical simulations in Section IV reveals that they are more accurate than the frozen spin approximation.

\section{Model and Perturbative Solution}

In this work, we describe the dynamics of the spin model which is generated by the Hamiltonian ($\hbar=1$)  \citep{law1,vardi},

\begin{equation}
H= 2\kappa \hat{J}_{z}^{2}+\Omega \hat{J}_{x}.
\end{equation}

For instance, $H$ can be generated by a trapped spinor Bose condensate with two populated hyperfine spin states $|a\rangle$ and $|b\rangle$  interacting with an external radio-frequency or microwave field. As a concrete example, the two hyperfine levels could be  the levels $|F=2,m_{f}=1\rangle$ and $|F=1,m_{f}=-1\rangle$ of $^{87}{\text {Rb}}$  atoms.  Provided that $\hat{a}_{1}$ and $\hat{a}_{2}$ denote the atomic annihilation operators corresponding to the hyperfine levels $|a\rangle$ and $|b\rangle$,  respectively, the angular momentum operators $\hat{J}_{+}=(\hat{J}_{-})^{\dagger}=\hat{a}_{2}^{\dagger}\hat{a}_{1}$, $\hat{J}_{z}=(\hat{a}_{2}^{\dagger}\hat{a}_{2}-\hat{a}_{1}^{\dagger}\hat{a}_{1})/2$ obey the $SU(2)$ Lie algebra. Furthermore, the Hamiltonian (1) commutes with the particle number operator $\hat{N}= \hat{a}_{2}^{\dagger}\hat{a}_{2}+\hat{a}_{1}^{\dagger}\hat{a}_{1} $, so that the total particle number is conserved.  Note that we neglect in (1) an additional term proportional to $\hat{J}_{z}$ by assuming equal intra-species interaction strength and the same trapping potential for both spin states \citep{chen}. The frequency $\Omega$ is controlled by the strength  of the external field and the  parameter $\kappa$, which describes the strength of the one-axis twisting, depends upon the inter- and intra-species two-body $s$-wave scattering lengths \citep{chen}. It turns out that the term $2 \kappa \hat{J}_{z}^{2}$ is essential for generating spin squeezing.

\subsection{Heisenberg Initial Value Problem}

We start with writing down the Heisenberg equations of motion for the respective time-dependent operators $\hat{J}_{x}(t)$, $\hat{J}_{y}(t)$ and $\hat{J}_{z}(t)$ in the Heisenberg picture:

\begin{equation}
\dot{\hat{J}}_{x}(t)=-2 \kappa [\hat{J}_{y}(t) \hat{J}_{z}(t)+\hat{J}_{z}(t) \hat{J}_{y}(t)],
\end{equation}

\begin{equation}
\dot{\hat{J}}_{y}(t)=-\Omega \hat{J}_{z}(t)+2 \kappa [\hat{J}_{z}(t) \hat{J}_{x}(t)+\hat{J}_{x}(t) \hat{J}_{z}(t)],
\end{equation}

\begin{equation}
\dot{\hat{J}}_{z}(t)=\Omega \hat{J}_{y}(t).
\end{equation}

In the following we aim at solving these Heisenberg equations of motion for general initial operators $\hat{J}_{i}(0)$ $(i=x,y,z)$ at time $t=0$. Later on, when expectation values are evaluated, we assume that the system is initially prepared in the lowest eigenstate $|J, m_{x}=-J\rangle$ of $\hat{J}_{x}(0)$, i.e., we have $\hat{J}_{x}(0) |J, m_{x}=-J\rangle=-J |J, m_{x}=-J\rangle$. Thus, the corresponding expectation values of the initial operators $\hat{J}_{i}(0)$ $i=x,y,z$ read $\langle \hat{J}_{x} (0) \rangle=-J$ and $\langle \hat{J}_{y} (0) \rangle=\langle \hat{J}_{z} (0) \rangle = 0$.

\noindent Combining Eqs.~(2) and (4), we find,

\begin{equation}
\dot{\hat{J}}_{x}(t)=- \frac{\kappa}{\Omega} \frac{d}{dt} \hat{J}_{z}^{2}(t),
\end{equation}

\noindent which has the following solution:

\begin{equation}
\hat{J}_{x}(t)=\hat{J}_{x}(0)+\frac{\kappa}{\Omega} \left[ \hat{J}_{z}^{2}(0) - \hat{J}_{z}^{2}(t) \right].
\end{equation}

\noindent Substituting this expression for $\hat{J}_{x}(t)$ into Eq.~(3) and using Eq.~(4) yields

\begin{equation}
\ddot{\hat{J}}_{z}(t)-2 i \kappa \dot{\hat{J}}_{z}(t)+\left[ \Omega^{2}-4 \kappa \Omega \hat{J}_{x}(0)\right] \hat{J}_{z}(t)= 4 \kappa^{2} \left[ \hat{J}_{z}^{2}(0)-\hat{J}_{z}^{2}(t)\right] \hat{J}_{z}(t).
\end{equation}

The second-order operator valued differential equation (7) is not exactly solvable due to its nonlinearity so we have to resort to approximative solution.  Therefore we review in Section II.B the commonly used frozen spin approximation, which was originally introduced in Ref. \citep{law1}. Then we work out in detail our strategy, where we treat both nonlinear terms on the right-hand side up to first order in perturbation theory. The zeroth-order solution $\hat{J}_{z}^{0}(t)$ is found in Section II.C by putting the right-hand side of Eq.~(7) to zero. Inserting $\hat{J}_{z}^{0}(t)$ then on the right-hand side yields two inhomogeneities. Thus solving Eq.~(7) in first order yields the corresponding corrections $\hat{J}_{z}^{I}(t)$ and $\hat{J}_{z}^{II}(t)$, which are determined in Section II.D and II.E, respectively.

\subsection{Frozen Spin Approximation}

Provided that $\Omega>>\kappa$, the external field forces the total spin to remain polarized in the direction of $ \langle \hat{J}_{x}(0) \rangle = -J$  as it costs energy to change the spin vector. Consequently, $\hat{J}_{x}$ remains approximately unchanged and one can replace $\hat{J}_{x}$ by $-J$ in the Heisenberg differential Eqs. (2)--(4). This so-called frozen spin approximation results in the operator valued differential equation

\begin{equation}
\ddot{\hat{J}}_{z}^{\text{fs}}(t)=-(\Omega^{2}+ 4 \kappa \Omega J) \hat{J}_{z}^{\text{fs}}(t).
\end{equation}

\noindent It is solved by

\begin{equation}
\hat{J}_{z}^{\text{fs}}(t)=\hat{J}_{z}(0) \cos{\omega_{\text{fs}} t}+ \frac{\Omega \hat{J}_{y}(0)}{\omega_{\text{fs}}} \sin{\omega_{\text{fs}} t},
\end{equation}

\noindent with the frozen spin frequency

\begin{equation}
\omega_{\text{fs}}=\sqrt{\Omega^{2}+ 4 \kappa \Omega J}.
\end{equation}

\noindent This result of the frozen spin approximation will later on be used as a reference to estimate the accuracy of our Heisenberg operator approach.

\noindent The corresponding expression for $\hat{J}_{y}^{\text{fs}}(t)$ is found by substituting (9) in (4) as

\begin{equation}
\hat{J}_{y}^{\text{fs}}(t)= -\frac{\omega_{\text{fs}} \hat{J}_{z}(0)}{\Omega} \sin{\omega_{\text{fs}} t}+ \hat{J}_{y}(0) \cos{\omega_{\text{fs}} t},
\end{equation}

\noindent This result will be utilized to determine the spin squeezing along the $y$-axis.

\subsection{Zeroth-Order Solution}

\noindent Now we turn to our solution strategy and determine at first the zeroth-order solution $\hat{J}_{z}^{(0)}(t)$ of the homogeneous equation

\begin{equation}
\ddot{\hat{J}}_{z}^{(0)}(t)-2 i \kappa \dot{\hat{J}}_{z}^{(0)}(t)+ \hat{\tilde{\omega}}^{2} \hat{J}_{z}^{(0)}(t)= 0,
\end{equation}

\noindent where we have introduced the abbreviation  $\hat {\tilde{\omega}}= \sqrt{\Omega^{2}-4 \kappa \Omega \hat{J}_{x}(0)}$. To this end we make the ansatz  $\hat{J}_{z}^{(0)}(t)= e^{\hat{K}t} \hat{O}$, where the auxiliary operators $\hat{K}$ and $\hat{O}$ are determined as follows. Inserting this ansatz in Eq.~(12) yields

\begin{equation}
\hat{K}^{2}-2 i \kappa \hat{K}+ \hat{\tilde{\omega}}^{2} \hat{K}= 0,
\end{equation}

\noindent which has the roots $\hat{K}_{1}=i (\kappa+\hat{\omega})$ and $\hat{K}_{2}=i (\kappa-\hat{\omega})$, with $\hat{\omega}=\sqrt{\kappa^{2}+\hat{\tilde{\omega}}^{2}}$. As Eq.~(12) is linear, the superposition principle yields the homogeneous solution

\begin{equation}
\hat{J}_{z}^{(0)}(t)=e^{\hat{K}_{1} t} \hat{O}_{1}+e^{\hat{K}_{2} t} \hat{O}_{2},
\end{equation}

\noindent whereas from Eq.~(4) we read off

\begin{equation}
\hat{J}_{y}^{(0)}(t)= \frac{\hat{K}_{1}}{\Omega} e^{\hat{K}_{1} t} \hat{O}_{1}+\frac{\hat{K}_{2}}{\Omega} e^{\hat{K}_{2} t} \hat{O}_{2}.
\end{equation}

\noindent By invoking the initial condition $\hat{J}_{y}^{(0)}(0)=\hat{J}_{y}(0)$, $\hat{J}_{z}^{(0)}(0)=\hat{J}_{z}(0)$, the operators $\hat{O}_{1}$, $\hat{O}_{2}$ are determined by the expressions

\begin{equation}
\hat{O}_{1}=\frac{\Omega i}{2 \hat{\omega}}\left[ \frac{i (\kappa-\hat{\omega})}{\Omega}\hat{J}_{z}(0)-\hat{J}_{y}(0)\right],
\end{equation}

\begin{equation}
\hat{O}_{2}=\frac{\Omega i}{2 \hat{\omega}}\left[ -\frac{i (\kappa+\hat{\omega})}{\Omega}\hat{J}_{z}(0)+\hat{J}_{y}(0)\right].
\end{equation}

\noindent Thus, the zeroth-order expression for  $\hat{J}_{z}(t)$ turns out to be

\begin{equation}
\hat{J}_{z}^{(0)}(t)=e^{i \kappa t} \left[ \left(\cos{\hat{\omega}t}-\frac{\kappa i}{\hat{\omega}}\sin{\hat{\omega}t} \right)\hat{J}_{z}(0) +\frac{\Omega}{\hat{\omega}} \sin{\hat{\omega}t} \hat{J}_{y}(0) \right].
\end{equation}

\noindent Equation (18) represents the exact zeroth-order operator solution for $\hat{J}_{z}(t)$, which can then be used with Eq.~(6) to evaluate the zeroth-order expression of the expectation value $\langle \hat{J}_{x}(t) \rangle $, yielding

\begin{equation}
\langle \hat{J}_{x}^{(0)}(t)\rangle=-J + \frac{\kappa J}{2 \Omega}- \frac{\kappa J}{2 \Omega} \left[ \left( \frac{\Omega}{\omega}-\frac{\kappa}{\omega}\right)^{2} \sin^{2}{\omega t}+\cos^{2}{\omega t}  \right].
\end{equation}

\noindent Note that the resulting frequency

\begin{equation}
\omega=\sqrt{\kappa^{2}+\Omega^{2}+4 \kappa \Omega (J-1)},
\end{equation}

\noindent differs from the one of the frozen spin approximation Eq.~(10) but in the limit $\Omega >> \kappa $,  $J>>1$ we read off from Eq.~(19) that we get $\langle \hat{J}_{x}^{(0)}(t) \rangle = -J = \langle \hat{J}_{x}^{\text{fs}}(t)\rangle$ and $\omega=\omega_{\text{fs}}$ i.e the frozen spin approximation follows from our zeroth order result as a special case.

\subsection{Particular and Homogeneous Solution I}

After having found the zeroth-order solution of Eq.~(7), we now turn our attention towards the first-order correction. The right-hand side of Eq.~(7) has two nonlinear terms, i.e. $4 \kappa^{2} \hat{J}_{z}^{2}(0)\hat{J}_{z}^{(0)}(t)$ and $-4 \kappa^{2} \hat{J}_{z}^{(0)  3}(t)$, which we treat now separately. At first, we determine the particular and the homogeneous solution of the differential equation

\begin{equation}
\ddot{\hat{J}}_{z}^{I}(t)-2 i \kappa \dot{\hat{J}}_{z}^{I}(t)+\left[ \Omega^{2}-4 \kappa \Omega \hat{J}_{x}(0)\right] \hat{J}_{z}^{I}(t)= 4 \kappa^{2}  \hat{J}_{z}^{2}(0) \hat{J}_{z}^{0}(t).
\end{equation}

\noindent Due to section II.C we recognize that the inhomogeneity of (21) oscillates with the same frequency $\hat{\omega}$ as its homogeneous part. Therefore, we perform for the particular solution  an ansatz which contains secular terms:

\begin{equation}
\hat{J}_{z,p}^{I}(t)=e^{i \kappa t} \left[ \hat{J}_{z}^{2}(0) \left(\hat{a}t \sin{\hat{\omega}t}+\hat{b}t \cos{\hat{\omega}t} \right)\hat{J}_{z}(0)+ \hat{J}_{z}^{2}(0) \left( \hat{c}t \sin{\hat{\omega}t}+\hat{d}t \cos{\hat{\omega}t} \right)\hat{J}_{y}(0) \right].
\end{equation}

\noindent Here $\hat{a}$, $\hat{b}$, $\hat{c}$, $\hat{d}$ denote operators, which  can be straight-forwardly determined by substituting Eq.~(22) into Eq.~(21) and by comparing the operator coefficients of the oscillating terms $\cos{\hat{\omega}t}$ and $\sin{\hat{\omega}t}$ on both sides of the resulting equation. This yields the particular solution

\begin{equation}
\hat{J}_{z,p}^{I}(t)=e^{i \kappa t} \left[ \hat{J}_{z}^{2}(0) \left(\frac{2 \kappa^{2}}{\hat{\omega}} t \sin{\hat{\omega}t}+\frac{2i \kappa^{3}}{\hat{\omega}^{2}} t \cos{\hat{\omega}t} \right) \hat{J}_{z}(0)-\hat{J}_{z}^{2}(0) \frac{2 \kappa^{2} \Omega}{\hat{\omega}^{2}} t \cos{\hat{\omega}t} \hat{J}_{y}(0) \right].
\end{equation}

\noindent Afterwards, we obtain the homogeneous solution of Eq.~(21) which has the form

\begin{equation}
\hat{J}_{z,h}^{I}(t)=e^{\hat{K}_{1}t}\hat{O}_{3}+e^{\hat{K}_{2}t}\hat{O}_{4}.
\end{equation}

\noindent Here $\hat{O}_{3}$ and $\hat{O}_{4}$ are unknown operators, which are determined from the initial conditions  $\hat{J}_{z,h}^{I}(0)=-\hat{J}_{z,p}^{I}(0)=0$, $\dot{\hat{J}}_{z,h}^{I}(0)=-\dot{\hat{J}}_{z,p}^{I}(0)$, yielding

\begin{equation}
\hat{J}_{z,h}^{I}(t)= \left( e^{\hat{K}_{1}t}-e^{\hat{K}_{2}t}\right) \frac{\hat{J}_{z}^{2}(0)\kappa^{2}}{i \hat{\omega}^{3}}  \left[i \kappa \hat{J}_{z}(0)-\Omega \hat{J}_{y}(0) \right].
\end{equation}

The secular terms $t \sin{\hat{\omega}t}$ and $t \cos{\hat{\omega}t}$ in Eq.~(23) seem to indicate that the solution $\hat{J}_{z}(t)$ grows unlimited in time. This finding contradicts, however, an exact numerical solution of the time-dependent Schr\"odinger equation governed by the Hamiltonian (1). Therefore, we follow Refs. \citep{mic,min,bog,axel,axel2,axel3} and introduce in the sum $\hat{J}_{z}^{0}(t)+\hat{J}_{z,h}^{I}(t)+\hat{J}_{z,p}^{I}(t)$ an effective frequency via $\hat{\omega}=\hat{\omega}_{\text{eff}}+ \kappa^{2} \hat{\omega}_{1}$, where $\hat{\omega}_{1}$ is determined by eliminating the secular terms up to first order in $\kappa^{2}$. This yields $\hat{\omega}_{1}= \hat{J}_{z}^{2}(0) \frac{2}{\hat{\omega}_{\text{eff}}} $ and the resulting bounded solution reads

\begin{eqnarray}
\hat{J}_{z}^{(0)}(t)+ \hat{J}_{z,h}^{I}(t)+ \hat{J}_{z,p}^{I}(t)&=& e^{i \kappa t}  \left[ \cos{\hat{\omega}_{\text{eff}}t}-\frac{i \kappa}{\hat{\omega}_{\text{eff}}} \sin{\hat{\omega}_{\text{eff}}t}+ 2i \kappa^{3} \hat{J}_{z}^{2}(0) \frac{\sin{\hat{\omega}_{\text{eff}}t}}{\hat{\omega}_{\text{eff}}^{3}} \right] \hat{J}_{z}(0)  \nonumber \\
&&+ e^{i \kappa t}  \left[ \frac{\Omega}{\hat{\omega}_{\text{eff}}} \sin{\hat{\omega}_{\text{eff}}t}-2 \hat{J}_{z}^{2}(0) \Omega \kappa^{2} \frac{\sin{\hat{\omega}_{\text{eff}}t}}{\hat{\omega}_{\text{eff}}^{3}} \right] \hat{J}_{y}(0) .
\end{eqnarray}

\noindent Here the effective frequency reads up to first order in $\kappa^{2}$ as follows:

\begin{equation}
\hat{\omega}_{\text{eff}}=\sqrt{\kappa^2+\Omega^{2}-4 \kappa \Omega \hat{J}_{x}(0)}-\frac{2 \kappa^{2} }{\sqrt{\kappa^2+\Omega^{2}-4 \kappa \Omega \hat{J}_{x}(0)}}\hat{J}_{z}^{2}(0).
\end{equation}

\subsection{Particular and Homogeneous Solution II}

Now it remains to solve the differential equation

\begin{equation}
\ddot{\hat{J}}_{z}^{II}(t)-2 i \kappa \dot{\hat{J}}_{z}^{II}(t)+\left[ \Omega^{2}-4 \kappa \Omega \hat{J}_{x}(0)\right] \hat{J}_{z}^{II}(t)= -4 \kappa^{2} \hat{J}_{z}^{(0)3},
\end{equation}

\noindent where $\hat{J}_{z}^{(0)}(t)$ follows from (26) by neglecting the $\kappa^{2}$ and $\kappa^{3}$ terms. In order to determine the particular solution of (28), we perform the ansatz

\begin{equation}
\hat{J}_{z,p}^{II}(t)= e^{i3\kappa t} \hat{F},
\end{equation}

\noindent with the abbreviation

\begin{eqnarray}
\hat{F}(t) &=& \hat{f}(t) \hat{J}_{z}(0)\hat{f}(t)\hat{J}_{z}(0)\hat{f}(t)\hat{J}_{z}(0)+\hat{g}(t) \hat{J}_{y}(0)\hat{g}(t) \hat{J}_{y}(0)\hat{g}(t) \hat{J}_{y}(0) \nonumber \\
&&+ \hat{f}(t) \hat{J}_{z}(0)\hat{f}(t) \hat{J}_{z}(0)\hat{g}(t) \hat{J}_{y}(0)+ \hat{f}(t) \hat{J}_{z}(0)\hat{g}(t) \hat{J}_{y}(0)\hat{f}(t) \hat{J}_{z}(0) \nonumber \\
&&+ \hat{f}(t) \hat{J}_{z}(0)\hat{g}(t) \hat{J}_{y}(0)\hat{g}(t) \hat{J}_{y}(0)+ \hat{g}(t) \hat{J}_{y}(0)\hat{f}(t) \hat{J}_{z}(0)\hat{f}(t) \hat{J}_{z}(0) \nonumber \\
&&+ \hat{g}(t) \hat{J}_{y}(0)\hat{f}(t) \hat{J}_{z}(0)\hat{g}(t) \hat{J}_{y}(0)+ \hat{g}(t) \hat{J}_{y}(0)\hat{g}(t) \hat{J}_{y}(0)\hat{f}(t) \hat{J}_{z}(0),
\end{eqnarray}

\bigskip

\noindent and the functions

\begin{equation}
\hat{f}(t)=\hat{\alpha}_{1} \cos{\hat{\omega}_{\text{eff}}t}+\hat{\beta}_{1} \sin{\hat{\omega}_{\text{eff}}t}, \hspace{1 cm} \hat{g}(t)=\hat{\alpha}_{2} \cos{\hat{\omega}_{\text{eff}}t}+\hat{\beta}_{2} \sin{\hat{\omega}_{\text{eff}}t}.
\end{equation}

\noindent Substituting (29)--(31) in (28) yields for the respective coefficients the result

\begin{equation}
\hat{\alpha}_{1}=\frac{28 \kappa^2}{25 \kappa^{2}-4 \hat{\omega}_{\text{eff}}^{2}}, \hat{\alpha}_{2}=\frac{i 8 \Omega \kappa}{25 \kappa^{2}-4 \hat{\omega}_{\text{eff}}^{2}}, \hat{\beta}_{1}=\frac{-4 i \kappa (5 \kappa^{2}+2 \hat{\omega}_{\text{eff}}^{2})}{\hat{\omega}_{\text{eff}}(25 \kappa^{2}-4 \hat{\omega}_{\text{eff}}^{2})}, \hat{\beta}_{2}=\frac{20 \kappa^{2} \Omega}{\hat{\omega}_{\text{eff}}(25 \kappa^{2}-4 \hat{\omega}_{\text{eff}}^{2})}.
\end{equation}

\bigskip

The corresponding homogeneous solution of (28) has the form

\begin{equation}
\hat{J}_{z,h}^{II}(t)=e^{\hat{K}_{1}t}\hat{O}_{5}+e^{\hat{K}_{2}t}\hat{O}_{6}.
\end{equation}

\noindent The initial conditions $\hat{J}_{z,h}^{II}(0)=-\hat{J}_{z,p}^{II}(0)=0$,  $\dot{\hat{J}}_{z,h}^{II}(0)=-\dot{\hat{J}}_{z,p}^{II}(0)$ yield $\hat{O}_{5}+\hat{O}_{6}=-\hat{F_{0}}$, $\hat{K}_{1} \hat{O}_{5}+\hat{K}_{2}\hat{O}_{6}=-\hat{G}_{0}$, where $\hat{F}_{0}=\hat{F}(t=0)$ and $\hat{G}_{0}=i 3 \kappa \hat{F}_{0}+\dot{\hat{F}}(t=0)$. Thus, the homogeneous solution reads

\begin{equation}
\hat{J}_{z,h}^{II}(t)=\frac{e^{i(\kappa+\hat{\omega}_{\text{eff}})t}}{2i \hat{\omega}_{\text{eff}}} [i(\kappa-\hat{\omega}_{\text{eff}}) \hat{F}_{0}-\hat{G}_{0}]+\frac{e^{i(\kappa-\hat{\omega}_{\text{eff}})t}}{2i \hat{\omega}_{\text{eff}}} [-i(\kappa+\hat{\omega}_{\text{eff}}) \hat{F}_{0}+\hat{G}_{0}].
\end{equation}

\section{Spin Squeezing}

The final expression for $\hat{J}_{z}(t)$ is the sum of the previously determined solutions (26), (29)--(32), (34), i.e.

\begin{equation}
\hat{J}_{z}(t)=\hat{J}_{z}^{(0)}(t)+\hat{J}_{z,h}^{I}(t)+\hat{J}_{z,p}^{I}(t)+\hat{J}_{z,h}^{II}(t)+\hat{J}_{z,p}^{II}(t).
\end{equation}

\noindent The expression (35) for $\hat{J}_{z}(t)$ is substituted into Eq.~(4) to evaluate the expression for $\hat{J}_{y}(t)$

\begin{eqnarray}
\hat{J}_{y}(t)&=& e^{i \kappa t}  \left[ \frac{ (\kappa^{2}-\hat{\omega}_{\text{eff}}^2)}{\Omega \hat{\omega}_{\text{eff}}} \sin{\hat{\omega}_{\text{eff}}t}+ 2i \kappa^{3} \hat{J}_{z}^{2}(0) \frac{\cos{\hat{\omega}_{\text{eff}}t}}{\Omega \hat{\omega}_{\text{eff}}^{2}}-2 \kappa^{4} \hat{J}_{z}^{2}(0) \frac{\sin{\hat{\omega}_{\text{eff}}t}}{\Omega \hat{\omega}_{\text{eff}}^{3}} \right] \hat{J}_{z}(0)  \nonumber \\
&&+ e^{i \kappa t}  \left[ \cos{\hat{\omega}_{\text{eff}}t}+\frac{i \kappa}{\hat{\omega}_{\text{eff}}} \sin{\hat{\omega}_{\text{eff}}t}-2 \hat{J}_{z}^{2}(0) \kappa^{2} \frac{\cos{\hat{\omega}_{\text{eff}}t}}{\hat{\omega}_{\text{eff}}^{2}}-2 i \hat{J}_{z}^{2}(0) \kappa^{3} \frac{\sin{\hat{\omega}_{\text{eff}}t}}{\hat{\omega}_{\text{eff}}^{3}} \right] \hat{J}_{y}(0) \nonumber \\
&&+\frac{i e^{i(\kappa+\hat{\omega}_{\text{eff}})t}(\kappa+\hat{\omega}_{\text{eff}})}{2i \Omega \hat{\omega}_{\text{eff}}} [i(\kappa-\hat{\omega}_{\text{eff}}) \hat{F}_{0}-\hat{G}_{0}]+\frac{i e^{i(\kappa-\hat{\omega}_{\text{eff}})t}(\kappa-\hat{\omega}_{\text{eff}})}{2i \Omega \hat{\omega}_{\text{eff}}} [-i(\kappa+\hat{\omega}_{\text{eff}}) \hat{F}_{0}+\hat{G}_{0}] \nonumber \\
&&+\frac{i3 \kappa e^{i3 \kappa t}}{\Omega}\hat{F}+\frac{e^{i3 \kappa t}}{\Omega} \dot{\hat{F}}
\end{eqnarray}


\noindent  Note that $\langle \hat{J}_{r}(t) \rangle =0 $. In a similar manner, the final expression (35) for $\hat{J}_{z}(t)$ is now substituted back into Eq.~(6) to evaluate the expectation value

\begin{equation}
\langle \hat{J}_{x} (t) \rangle= -J+ \frac{\kappa}{\Omega} \left[\frac{J}{2}-J_{1}(t)-J_{2}(t)-J_{3}(t)\right].
\end{equation}

\noindent The resulting expressions for $J_{1}(t),J_{2}(t), J_{3}(t)$ turn out to oscillate with the frequency

\begin{equation}
\omega_{\text{eff}}=\sqrt{\kappa^2+\Omega^{2}+4 \kappa \Omega (J-1)}-\frac{\kappa^{2} J}{\sqrt{\kappa^2+\Omega^{2}+4 \kappa \Omega (J-1)}}
\end{equation}

\noindent and are explicitly given in Appendix A.

Following the criteria of for spin squeezing of Kitagawa and Ueda in Ref. \citep{kita1}, we introduce the squeezing parameter

\begin{equation}
\xi_{s,\textbf{n}}=\frac{\sqrt{2} \langle (\Delta \hat{J}_{\textbf{n}})_{\text{min}}\rangle}{\sqrt{J}},
\end{equation}

\noindent where $\langle (\Delta \hat{J}_{\textbf{n}})_{\text{min}}\rangle$ is the smallest uncertainty of spin component $\hat{J}_{\textbf{n}}=\hat{\textbf{J}}.\textbf{n}$ perpendicular to the mean spin $\langle \hat{\textbf{J}} \rangle$. A state is said to be a squeezed-spin state provided that the inequality $\xi_{s,\textbf{n}}<1$ holds.
Since the mean spin points along the $x$-direction, the reduced spin fluctuations occur in the $yz$-plane.
The spin component normal to the mean spin is $\hat{\textbf{J}}_{\textbf{n}}=\hat{J}_{y} \sin{\theta}+\hat{J}_{z} \cos{\theta}$ \citep{jin1}. By minimizing the variance $(\Delta \hat{J}_{\textbf{n}})_{\text{min}}$ with respect to $\theta$, we find the squeezing angle as

\begin{equation}
\theta_{min}=\frac{1}{2}tan^{-1}(\frac{B}{A}),
\end{equation}

and the squeezing parameter

\begin{equation}
\xi_{s,\textbf{n}}=\frac{\sqrt{C-\sqrt{A^2+B^2}}}{\sqrt{J}}.
\end{equation}

Here $A=\langle \hat{J}_{z}^{2}-\hat{J}_{y}^{2} \rangle = L_{1}(t)+L_{2}(t)+L_{3}(t)-J_{1}(t)-J_{2}(t)-J_{3}(t)$, $B=\langle \hat{J}_{z} \hat{J}_{y}+\hat{J}_{y} \hat{J}_{z} \rangle = J [(\alpha'_{1}(t)-\beta'_{2}(t))F'(t)-(\alpha'_{2}(t)-\beta'_{1}(t))\cos {\omega_{\text{eff}} t}+\gamma'_{1}(t)G'(t)(2J-1)(3J+3)]$ and $C=\langle \hat{J}_{z}^{2}+\hat{J}_{y}^{2} \rangle=L_{1}(t)+L_{2}(t)+L_{3}(t)+J_{1}(t)+J_{2}(t)+J_{3}(t)$

\noindent The explicit expressions for $L_{1}(t)$, $L_{2}(t)$, $L_{3}(t)$,$\alpha'_{1}(t)$, $\alpha'_{2}(t)$, $\beta'_{1}(t)$, $\beta'_{2}(t)$,$\gamma'_{1}(t)$,$F'(t)$ and $G'(t)$ are given in Appendix A. We have used $\langle \hat{J}_{z}(t)\rangle=0$,  $\langle \hat{J}_{z}^{2}(0)\rangle$ $= J/2 $, $\langle \hat{J}_{y}(t)\rangle=0$,  $\langle \hat{J}_{y}^{2}(0)\rangle$ $= J/2 $ in determining $\xi_{s,\textbf{n}}$.

Note that the corresponding expression for the squeezing parameter under the frozen spin approximation is found from Eqns.~(9),(11) and (39) as

\begin{equation}
\xi_{s,\textbf{n}}^{\text{fs}} =\frac{\sqrt{C_{fs}-\sqrt{A_{fs}^2+B_{fs}^2}}}{\sqrt{J}},
\end{equation}

where

\begin{equation}
A_{fs}=\frac{J}{2} \left[ \left( \frac{\omega_{\text{fs}}^2}{\Omega^{2}}-\frac{\Omega^{2}}{\omega_{\text{fs}}^{2}}  \right) \right]\sin^{2}{\omega_{\text{fs}} t},
\end{equation}

\begin{equation}
B_{fs}= \frac{J (\Omega^{2}-\omega_{\text{fs}}^2)}{\Omega \omega_{\text{fs}}}\sin{\omega_{\text{fs}} t}\cos{\omega_{\text{fs}} t},
\end{equation}

\begin{equation}
C_{fs}= \frac{J}{2}\left[\left( \frac{\omega_{\text{fs}}^2}{\Omega^{2}}+\frac{\Omega^{2}}{\omega_{\text{fs}}^{2}}  \right) \sin^{2}{\omega_{\text{fs}} t}+2 \cos^{2}{\omega_{\text{fs}} t} \right].
\end{equation}

\noindent The expression of $\theta_{min}$ for the frozen spin approximation is the same as (40) with $A$ and $B$ replaced by $A_{fs}$ and $B_{fs}$ respectively. In analogy to Section II. C, $\xi_{s,\textbf{n}}^{\text{fs}}$ in (42) follow in the limit $\Omega>>\kappa$ and $J>>1$ from $\xi_{s,\textbf{n}}$ in (41).

\section{Numerical Solution and Results}

After having determined a perturbative solution in the previous section, we now describe the exact numerical solution of the time-dependent \text{Schr\"odinger} equation governed by the Hamiltonian (1). In our work we assume positive $\Omega$ and $\kappa$, the latter corresponding to a repulsive inter- and intra-species interaction.  The state vector at any time $t$ can be expanded as $|\psi(t)\rangle=\sum_{m=-J}^{+J} c_{m}(t) |J, m \rangle$. The corresponding amplitudes $c_{m}(t)$ obey the time-dependent Schr\"odinger equation

\begin{equation}
i \frac{d c_{m}(t)}{dt}= 2 \kappa m^{2} c_{m}(t) +\zeta_{m} c_{m-1}(t) +\zeta_{-m} c_{m+1}(t),
\end{equation}

\noindent where we have introduced $\zeta_{m}=\frac{\Omega}{2}\sqrt{(J+m)(J-m+1)}$ with $\zeta_{-J}=0$ and $\zeta_{\pm m}=\zeta_{\mp m+1}$. We consider that the spin system starts from the lowest eigenstate $|J, m_{x}=-J\rangle$ of $\hat{J}_{x}$,  i.e., $\hat{J}_{x} |J, m_{x}=-J\rangle=-J |J, m_{x}=-J\rangle$. The resulting amplitudes of the initial state read

\begin{equation}
c_{m}(0)= \frac{(-1)^{J+m}}{2^{J}} \sqrt{\frac{(2 J)!}{(J-m)!(J+m)!}}
\end{equation}

\noindent and satisfy $c_{-m}(0)=c_{m}(0)$ for total number of atoms, i.e. $N=2 J$ and $c_{-m}(0)=- c_{m}(0)$ for odd total number of particles. This yields initially the expectation value $\langle \hat{J}_{z}(0)\rangle =0 $ and the variance $\langle \hat{J}_{z}^{2}(0)\rangle = J/2$. The symmetry properties of $\zeta_{\pm m}$ and $c_{m}(0)$ lead to $c_{-m}(t)=\pm c_{m}(t)$ and to the time dependent expectation values $\langle \hat{J}_{y}(t) \rangle = \langle \hat{J}_{z}(t) \rangle = 0$ as well as $\langle \hat{J}_{x}(t) \rangle \neq 0 $. This implies that the mean spin always points along the $x$-axis.

\begin{figure}[t]
\hspace{-0.0cm}
\begin{tabular}{cc}
\includegraphics [scale=0.60]{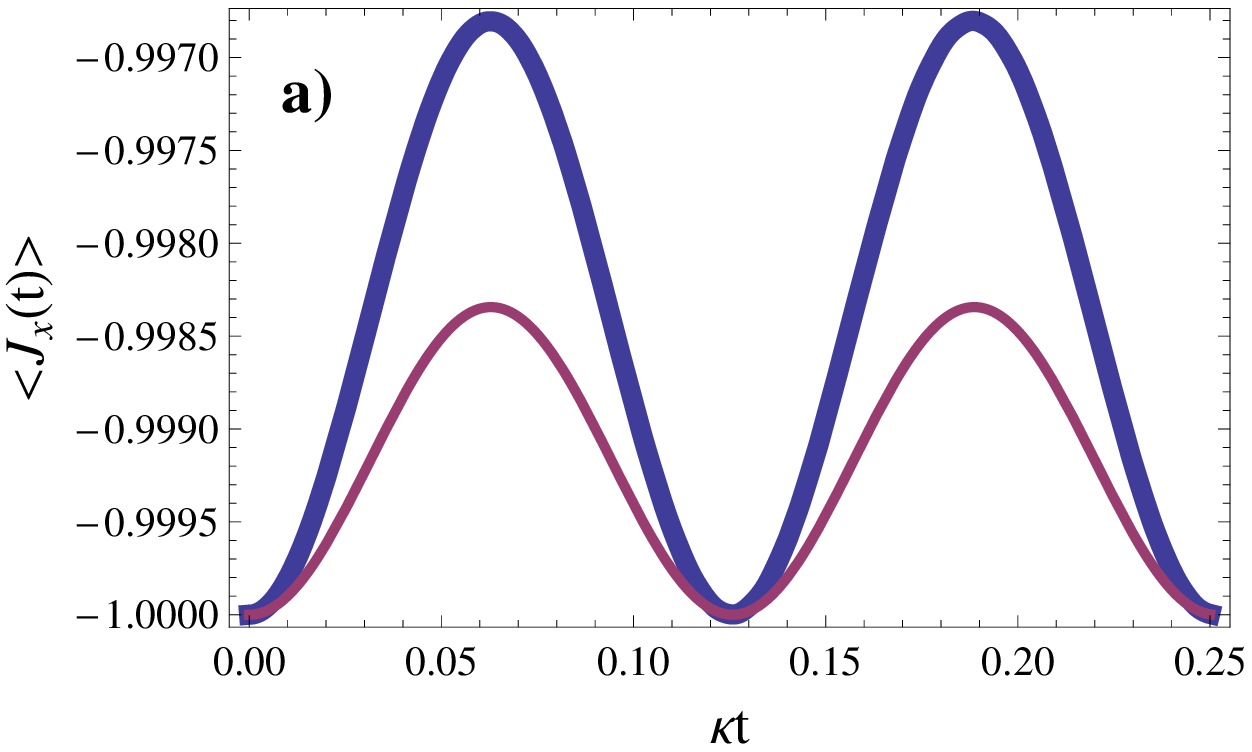}& \includegraphics [scale=0.60]{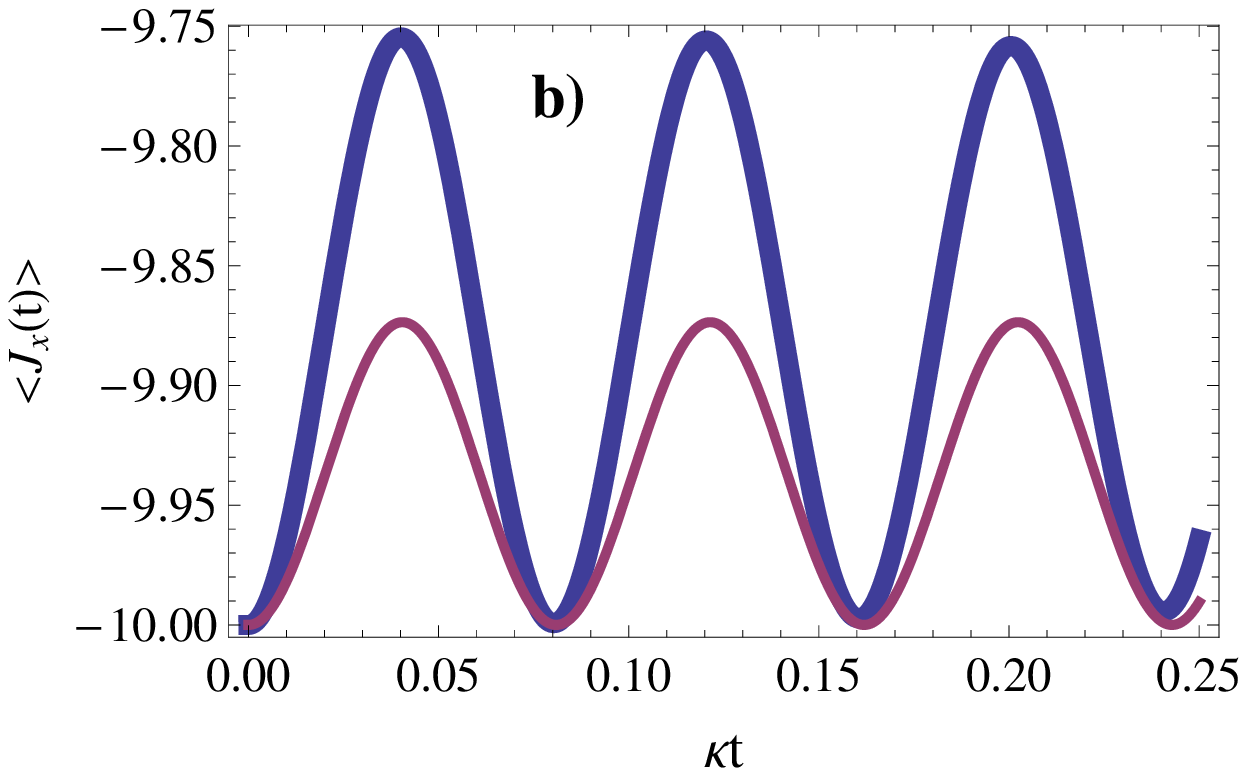}\\
\end{tabular}
\caption{(color online) Expectation value $\langle \hat{J}_{x} (t) \rangle$ (a) for $J=1$ and (b) $J=10$ with $\Omega/\kappa=25$ as a function of time in units of $\kappa$. Thick blue line depicts the numerics and thin red line represents the perturbative result (37).  }
\end{figure}\label{fig1}

\begin{figure}[t]
\hspace{-0.0cm}
\begin{tabular}{cc}
\includegraphics [scale=0.60]{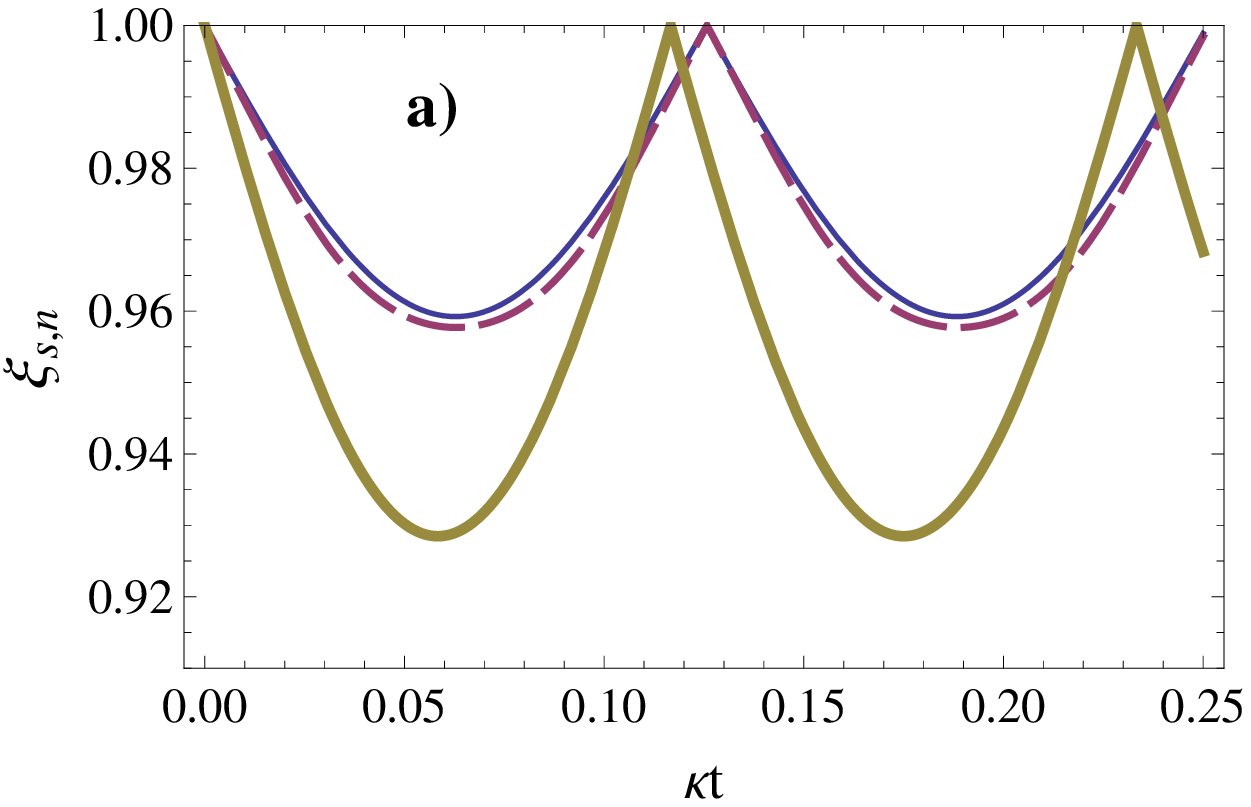}& \includegraphics [scale=0.60]{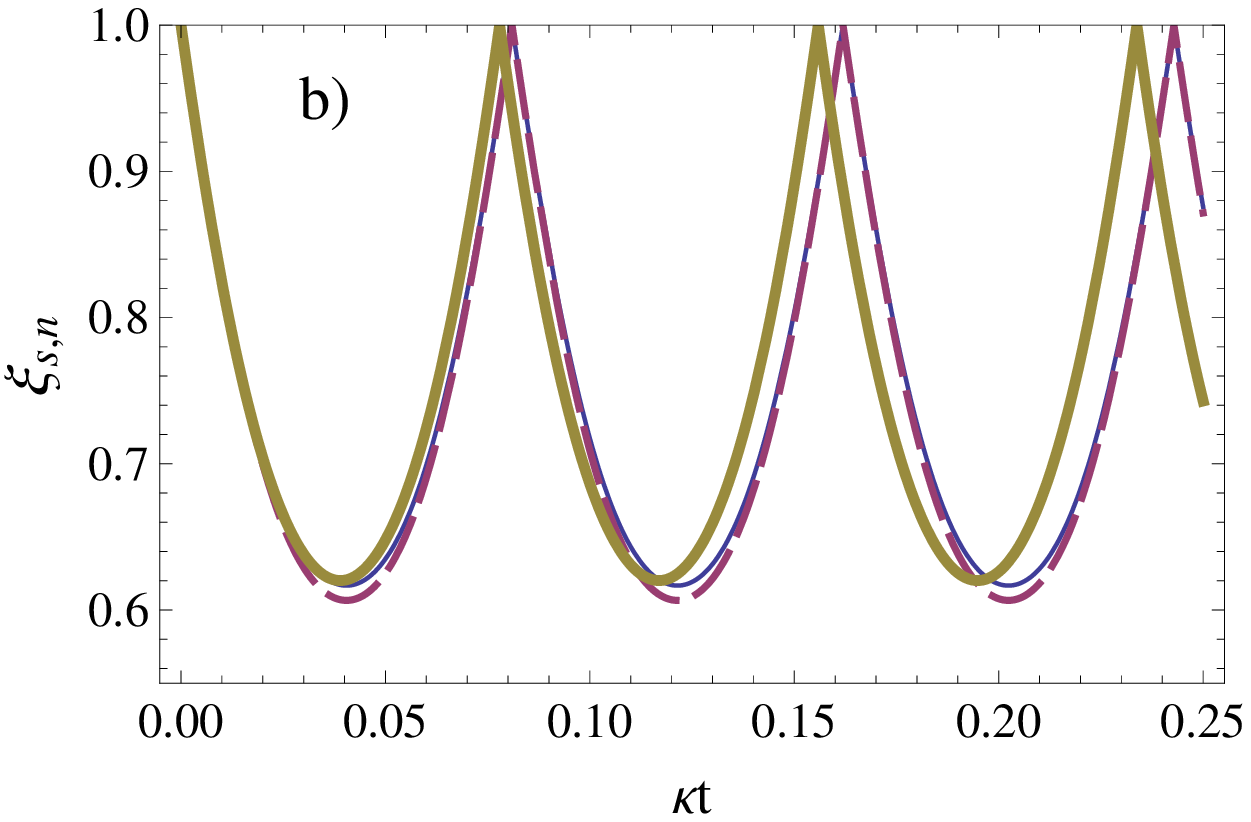}\\
\end{tabular}
\caption{(color online) Squeezing parameter (a) for $J=1$ and (b) $J=10$ with $\Omega/\kappa=25$ as a function of time in units of $\kappa$. Thin line denotes the numerics, thin dashed line stands for the perturbative corrected result (41), and thick line is the frozen spin approximation (42).  }
\end{figure}\label{fig2}

\begin{figure}[t]
\hspace{-0.0cm}
\includegraphics [scale=0.80]{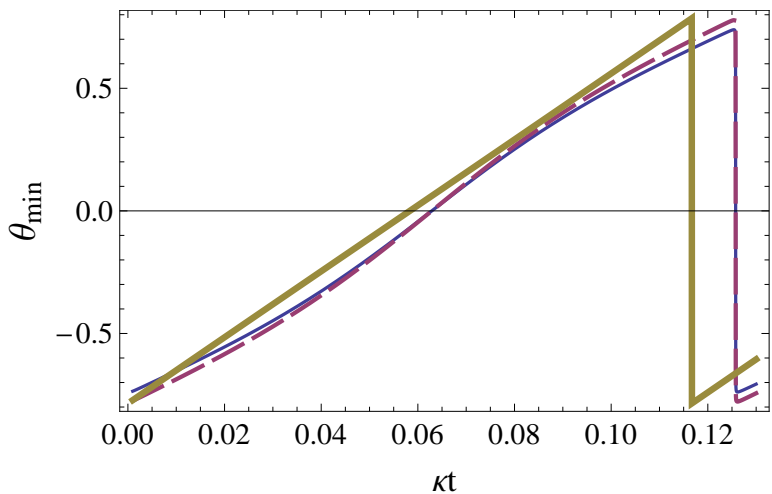}\\
\caption{(color online) Plot of $\theta_{min}$ for $J=1$ with $\Omega/\kappa=25$ as a function of time in units of $\kappa$. Thin line denotes the numerics, thin dashed line stands for the perturbative corrected result, and thick line is the frozen spin approximation.  }
\end{figure}\label{fig3}

In Fig. 1, we compare the expression for $\langle \hat{J}_{x} (t) \rangle$ as a function of dimensionless time obtained from our perturbative result of Eq. (37) with that obtained from exact numerics. We observe that the numerics and the perturbative corrected result turn out to agree better for smaller $J$.

In Fig. 2, we compare the results for the squeezing parameter along the direction perpendicular to the mean spin direction obtained from the Heisenberg operator method (41), exact numerics and that obtained from frozen spin approximation (42) for the two values of $J=1$ and $J=10$. Notably, the operator method result matches very well with the numerical data. For $J=1$, the frozen spin approximation result differs significantly from both the numerics and the Heisenberg operator method results, while for $J=10$ the match is much better. Thus, the perturbative result matches almost exactly with the numerics for both small and large $J$. For large $J$, the frozen spin approximation result approaches the numerical result. The results obtained in Fig. 2 illustrate the accuracy and effectiveness of our analytical perturbative operator method.

\begin{figure}[t]
\hspace{-0.0cm}
\begin{tabular}{cc}
\includegraphics [scale=0.60]{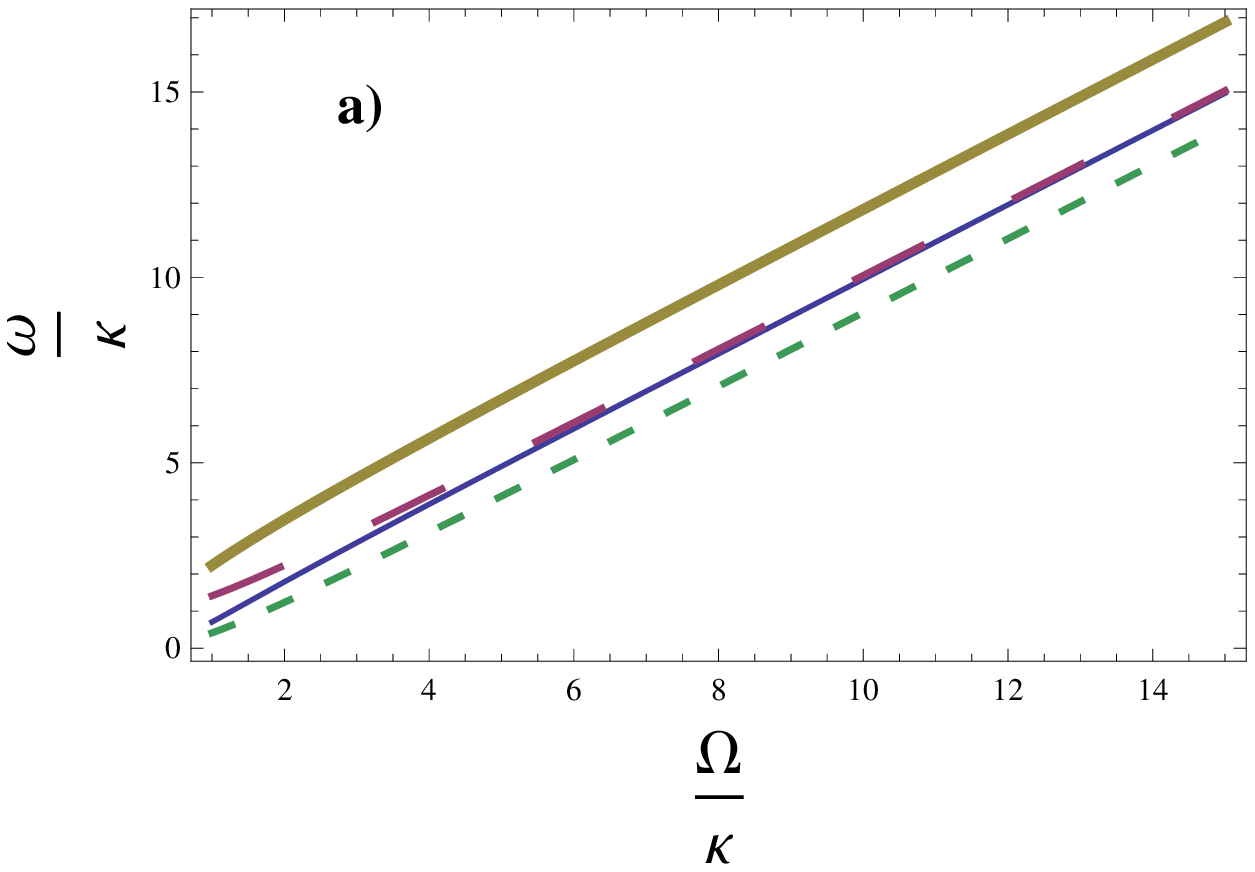}& \includegraphics [scale=0.60] {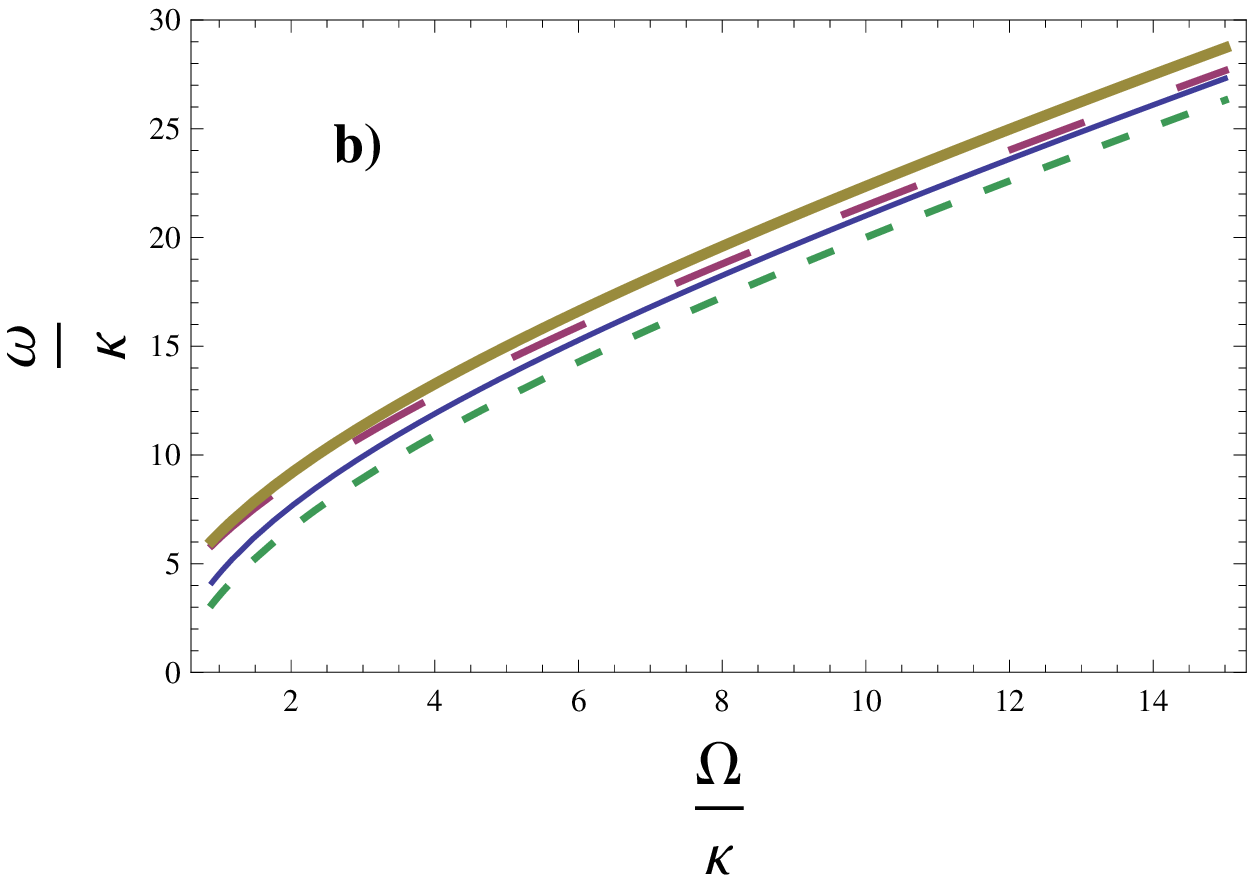}\\
\end{tabular}
\caption{(color online) Frequency of $\langle \hat{J}_{x}(t)\rangle$ in the units of  $\kappa$ as a function of $\Omega/\kappa$ for  $J=1$ (Plot a) and $J=10$ (Plot b). Thick line is the frozen spin approximation (10), long-dashed purple line is the zeroth order result (20), thin blue line is the perturbative corrected result (37) and short-dashed blue line is the numerics.  }
\end{figure}\label{fig4}

Fig.3 displays the comparative plot of $\theta_{min}$ as a function of time obtained from the Heisenberg operator method, exact numerics and that obtained from frozen spin approximation for $J=1$. As before, the perturbative result matches better with the numerical result. The value of $\theta_{min}$ when the squeezing parameter is unity is $\pm 0.78$ while its value is $0$ when the squeezing parameter is minimal.

In order to compare further the various results, we have plotted the frequency of $\langle \hat{J}_{x}(t)\rangle$ as a function of $\Omega/\kappa$ in Fig. 4. As evident from Fig. 4, the perturbative corrected result (37) is closest to the numerics while the frozen spin approximation (10) deviates much from the numerical result. For large $\Omega$ it turns out that the zeroth order (20) and the perturbative corrected result (37) match, while for low $\Omega$, the perturbative result (37) matches better with the numerics.

\section{Conclusions}

The expression for the spin squeezing parameter is evaluated by solving the one-axis twisting Hamiltonian (1) using the quantum mechanical perturbative operator method. We have demonstrated that the results obtained from the Heisenberg operator method coincide much better with that obtained from numerical results as compared to the frozen spin approximation results \citep{law1}. This finding nourishes the prospect that the Heisenberg operator method might turn out to be useful for analyzing also other spin squeezing dynamics.

\begin{acknowledgments}
A. Bhattacherjee acknowledges financial support from the University Grants Commission, New Delhi under the UGC-Faculty Recharge Programme and the German Academic Exchange Service (DAAD) fellowship to pursue joint research at the Technical University of Kaiserslautern, Germany. A. Bhattacherjee is also grateful to Sebastian Eggert for giving the opportunity to work in his group at the Technical University of Kaiserslautern, Germany. D. Sharma acknowledges financial support from the Jawaharlal Nehru University under the university scholarship scheme. This work was also supported by the German Research Foundation (DFG) via the Collaborative
Research Centers SFB/TR 49 and, at the final stage, SFB/TR 185.
\end{acknowledgments}

\section{Author contribution statement}
A.B.B. and A.P. conceived the methodology used in this model. A.B.B. did the calculations while D.S. did the numerical simulations. A.P. and A.B.B. analyzed
the graphs, discussed the results and prepared the manuscript.

\section{Appendix A}

\noindent Here we list the expressions, which appear in the expectation value (36) and the squeezing parameter in (41):

\begin{equation}
J_{1}(t) = \frac{J}{2} \left \{ \cos^{2}{{\omega}_{\text{eff}}t} + \left[  \frac{(\Omega-\kappa)^{2}}{{\omega}_{\text{eff}}^{2}} \left( 1-\frac{(3J-1)\kappa^{2}}{{\omega}_{\text{eff}}^{2}} \right) ^{2} + \frac{(2J-1)(3J+3)(\Omega-\kappa)^{2}\kappa^{4}}{{\omega}_{\text{eff}}^{6}}  \right] \sin^{2}{{\omega}_{\text{eff}} t}  \right \},
\end{equation}

\begin{eqnarray}
J_{2}(t) &=& \frac{J}{8} \{ \left[-f_{1}^{3}(3 J - 1) + g_{1}^{3}(J - 1) - f_{1} g_{1}^{2} (3 J - 1)\right]^{2} \nonumber \\
&+& \left[(f_{1}^{3} - g_{1}^{3} - 3 f_{1} g_{1}^{2})^{2} \right] (2 J - 1) (3 J + 3)+f_{1}^{4} g_{1}^{2} \left[(7 J - 3)^{2} + 9 (2 J - 1)(3 J + 3) \right] \} ,
\end{eqnarray}

\begin{equation}
J_{3}(t)=  F_{1} \cos^{2}{\omega_{\text{eff}} t} + G_{1}  \sin^{2}{\omega_{\text{eff}} t},
\end{equation}

\begin{eqnarray}
L_{1}(t) &=& \frac{J}{2} \left \{ \left [ 1+ (\frac{\kappa^{6}}{\Omega^{2}}+\kappa^{4})\frac{15J^{2}-3J-2}{{\omega}_{\text{eff}}^{4}} \right]\cos^{2}{{\omega}_{\text{eff}}t}\right \} \nonumber \\
&+& \frac{J}{2} \left \{ \left[  \frac{(\kappa^{2}-{\omega}_{\text{eff}}^{2})^{2}}{{\omega}_{\text{eff}}^{2} \Omega^{2}}+\frac{\kappa^{2}}{\Omega^{2}}+ \kappa^{6} \frac{(15J^{2}-3J-2)}{{\omega}_{\text{eff}}^{6}}  \right] \sin^{2}{{\omega}_{\text{eff}} t}  \right \},
\end{eqnarray}

\begin{equation}
L_{2}(t)= \frac{9 \kappa^{2}}{\Omega^{2}}J_{2}(t),
\end{equation}

\begin{equation}
L_{3}(t)= \left [ \frac{({\omega}_{\text{eff}}^{2}-\kappa^{2})^{2}}{{\omega}_{\text{eff}}^{2} \Omega^{2}}F_{1} +\frac{9 \kappa^{4}}{{\omega}_{\text{eff}}^{2} \Omega^{2}}F_{1}+\frac{{\omega}_{\text{eff}}^{2} }{\Omega^{2}} G_{1}  \right ] \sin^{2}{\omega_{\text{eff}} t}+ \left [  \frac{9 \kappa^{2}}{\Omega^{2}} F_{1}+\frac{\kappa^{2}}{\Omega^{2}}G_{1} \right ] \cos^{2}{\omega_{\text{eff}} t},
\end{equation}

\begin{equation}
\alpha'_{1}(t)= \left[ 1- \frac{\kappa^{2}(3J-1)}{\omega_{\text{eff}}^2} \right]\cos {\omega_{\text{eff}} t},
\end{equation}

\begin{equation}
\alpha'_{2}(t)= \left[ \frac{\kappa}{\Omega} -\frac{\kappa^{3}(3J-1)}{\omega_{\text{eff}}^3} \right]\sin {\omega_{\text{eff}} t},
\end{equation}

\begin{equation}
\beta'_{1}(t)= \left[ \frac{(\kappa^2-\omega_{\text{eff}}^2)}{\Omega \omega_{\text{eff}}} - \frac{\kappa^{4}(3J-1)}{\Omega \omega_{\text{eff}}^3}  \right]\sin {\omega_{\text{eff}} t},
\end{equation}

\begin{equation}
\beta'_{2}(t)= \frac{\kappa^{3} (3J-1)}{\Omega \omega_{\text{eff}}^2}\cos {\omega_{\text{eff}} t},
\end{equation}

\begin{equation}
\gamma'_{1}= \frac{\kappa^{2}}{\omega_{\text{eff}}^2} \left( 1+ \frac{\kappa}{\Omega} \right)\cos {\omega_{\text{eff}} t},
\end{equation}

\begin{equation}
F'(t)= \left[ \frac{\Omega-\kappa}{\omega_{\text{eff}}} + \frac{\kappa^{2}(3J-1)(\kappa-\Omega)}{\omega_{\text{eff}}^{3}} \right]\sin {\omega_{\text{eff}} t},
\end{equation}

\begin{equation}
G'(t)=\frac{\kappa^{2}(\kappa-\Omega)}{\omega_{\text{eff}}^{3}}\sin {\omega_{\text{eff}} t}.
\end{equation}

\bigskip

\noindent The respective abbreviations in (48),(49),(50),(52) and (53) are given by

\begin{equation}
f_{1}= \sqrt{\alpha_{1}^{2} \cos^{2}{\omega_{\text{eff}} t} + \beta_{1}^{2} \sin^{2}{\omega_{\text{eff}}t}},
\end{equation}

\begin{equation}
g_{1}= \sqrt{\alpha_{2}^{2} \cos^{2}{\omega_{\text{eff}} t} + \beta_{2}^{2} \sin^{2}{\omega_{\text{eff}}t}},
\end{equation}

\begin{eqnarray}
F_{1} &=& \frac{J}{8} \left[-\alpha_{1}^{3}(3 J - 1) + \alpha_{2}^{3}(J - 1) - \alpha_{1} \alpha_{2}^{2} (3 J - 1)\right]^{2} \nonumber \\
&+& (\alpha_{1}^{3} - \alpha_{2}^{3} - 3 \alpha_{1} \alpha_{2}^{2})^{2} (2 J - 1) (3 J + 3)+\frac{J}{8} \alpha_{1}^{4} \alpha_{2}^{2}\left[(7 J - 3)^{2} + 9 (2 J - 1)(3 J + 3)\right] ,
\end{eqnarray}

\begin{eqnarray}
G_{1} &=& \frac{J}{8} (4\alpha_{1}^{3} \beta_{1} \alpha_{2}^{2} + 2 \alpha_{1}^{4} \alpha_{2} \beta_{2}) \left[(7 J - 3)^{2} - 9 (2 J - 1) (3  J + 3) \right]  \nonumber \\
&+& \frac{J}{8} \left[-3 \alpha_{1}^{2} \beta_{1}(3 J - 1) + 3 \alpha_{2}^{2} \beta_{2} (J - 1) - (\beta_{1} \alpha_{2}^{2} + 2 \alpha_{1} \alpha_{2} \beta_{2})(3 J - 1) \right]^2  \nonumber \\
&+& (3 \alpha_{1}^{2} \beta_{1} - 3 \alpha_{2}^{2} \beta_{2} - 3 \beta_{1} \alpha_{2}^{2} - 6 \alpha_{1} \alpha_{2} \beta_{2})^2 (2 J - 1) (3 J + 3),
\end{eqnarray}

\bigskip

\noindent where we have

\begin{equation}
\alpha_{1} = \frac{28 \kappa^{2}}{25 \kappa^{2}-4 \omega_{\text{eff}}^{2}}, \alpha_{2} = \frac{8 \Omega \kappa}{25 \kappa^{2} - 4 \omega_{\text{eff}}^{2}}, \beta_{1} = \frac{-4 \kappa(5 \kappa^{2} + 2 \omega_{\text{eff}}^{2})}{ \omega_{\text{eff}} (25 \kappa^{2} - 4 \omega_{\text{eff}}^{2})}, \beta_{2} = \frac{20 \Omega \kappa^{2}}{ \omega_{\text{eff}} (25 \kappa^{2} - 4 \omega_{\text{eff}}^{2})}.
\end{equation}


\begin{thebibliography}{99}

\bibitem{gross}
C. Gross, J. Phys. B: At. Mol. Opt. Phys. \textbf{45}, 103001 (2012).
%
\bibitem{ma}
J. Ma, X. Wang, C. P. Sun, and F. Nori, Phys. Rep. \textbf{509}, 89  (2011).
%
\bibitem{wine1}
D. J. Wineland, J. J. Bollinger, W. M. Itano, F. L. Moore, and D. J. Heinzen, Phys. Rev. A \textbf{46}, R6797  (1992).
%
\bibitem{wine2}
D. J. Wineland, J.J. Bollinger, W. M. Itano, and D. J. Heinzen, Phys. Rev. A \textbf{50}, 67  (1994).
%
\bibitem{polzik}
E. S. Polzik, Nature \textbf{453}, 45 (2008).
%
\bibitem{cronin}
A. D. Cronin, J. Schmiedmayer, and D. E. Pritchard, Rev. Mod. Phys. \textbf{81}, 1051 (2009).
%
\bibitem{sorensen}
A. Sorensen, L. M. Duan, J. I. Cirac, and P. Zoller, Nature \textbf{409}, 63 (2001).
%
\bibitem{bigelow}
N. Bigelow, Nature \textbf{409}, 27 (2001).
%
\bibitem{gue}
O. Guehne and G. Toth, Phys. Rep. \textbf{474}, 1 (2009).
%
\bibitem{kita1}
M. Kitagawa and M. Ueda, Phys. Rev. A \textbf{47}, 5138  (1993).
%
\bibitem{tur}
Q. A. Turchette, C. S. Wood, B. E. King, C. J. Myatt, D. Leibfried, W. M. Itano, C. Monroe, and D. J. Wineland, Phys. Rev. Lett. \textbf{81}, 3631 (1998).
%
\bibitem{meyer}
V. Meyer, M. A. Rowe, D. Kielpinski, C. A. Sackett, W. M. Itano, C. Monroe, and D. J. Wineland, Phys. Rev. Lett. \textbf{86}, 5870 (2001).
%
\bibitem{leib}
D. Leibfried, M. D. Barrett, T. Schaetz, J. Britton, J. Chiaverini, W. M. Itano, J. D. Jost, C. Langer, D. J. Wineland, Science \textbf{304}, 1476 (2004).
%
\bibitem{wang}
X. Wang and B. C. Sanders, Phys. Rev. A \textbf{68}, 012101 (2003).
%
\bibitem{kor}
J. K. Korbicz, J. I. Cirac, and M. Lewenstein, Phys. Rev. Lett. \textbf{95}, 120502 (2005).
%
\bibitem{yi}
S. Yi and H. Pu, Phys. Rev. A \textbf{73}, 023602 (2006).
%
\bibitem{orz}
C. Orzel, A. K. Tuchman, M. L. Fenselau, M. Yasuda, and M. A. Kasevich, Science \textbf{291},  2386 (2001).
%
\bibitem{ried}
M. F. Riedel, Nature \textbf{464}, 1170 (2010).
%
\bibitem{est}
J. Esteve, C. Gross , A. Weller, S. Giovanazzi, and M. K. Oberthaler,  Nature \textbf{455}, 1216 (2008).
%
\bibitem{mue}
W. Muessel, H. Strobel, D. Linnemann, T. Zibold, B. Juli$\acute{a}$-Diaz, and M. K. Oberthaler, Phys. Rev A \textbf{92}, 023603 (2015).
%
\bibitem{Jaa}
M. Jaaskelainen and P. Meystre, Phys. Rev. A \textbf{73}, 013602 (2006).
%
\bibitem{law1}
C. K. Law, H. T. Ng, and P. T. Leung, Phys. Rev. A \textbf{63}, 055601 (2001).
%
\bibitem{jin1}
G. R. Jin and S. W. Kim, Phys. Rev. A \textbf{76}, 043621 (2007).
%
\bibitem{jin2}
G. R. Jin and C. K. Law, Phys. Rev. A \textbf{78}, 063620 (2008).
%
\bibitem{jin3}
G. R. Jin, X. W. Wang, and Y. W. Lu, J. Phys. B: At. Mol. Opt. Phys. \textbf{43}, 045301 (2010).
%
\bibitem{bhatt1}
A. B. Bhattacherjee, V. Ranjan, and M. Mohan, Int. J. Mod. Phys. B \textbf{17}, 2579 (2003).
%
\bibitem{fsa2}
Z. W. Bian and· X. B. Lai, Int. J. Theo. Phys. \textbf{52}, 3922 (2013).
%
\bibitem{fsa1}
G. J. Hu and X. X. Hu, Int. J. Theo. Phys. \textbf{53}, 533  (2014).
%
\bibitem{fsa3}
S. S. Li, H.G. Yi and R. H. Chen, Int. J. Theo. Phys. \textbf{52}, 1175 (2013).
%
\bibitem{fsa4}
J. Vidal, G. Palacios, and C. Aslangul, Phys. Rev. A \textbf{70},  062304 (2004).
%
\bibitem{fsa5}
Y. H. Jiang and S. S. Li, Int. J. Theo. Phys. \textbf{52}, 2826  (2013).
%
\bibitem{fsa6}
D. Kajtoch and E. Witkowska, Phys. Rev A \textbf{93}, 023627 (2016).
%
\bibitem{vardi}
A. Vardi and J. R. Anglin, Phys. Rev. Lett. \textbf{86}, 568 (2001).
%
\bibitem{chen}
G. Chen, X. Wang, J. Q. Liang, and Z. D. Wang, Phys. Rev. A \textbf{78}, 023634 (2008).
%
\bibitem{bog}
N. N.  Bogoliubov,  and Y. A.  Mitropolsky, Asymptotic Methods in the Theory of Non-Linear Oscillations, Gordon and Breach, New York, 1961.
%
\bibitem{min}
N. Minorsky, Nonlinear Oscillation, Van Nostrand, Princeton, 1962.
%
\bibitem{mic}
R.  Mickens, Introduction  to  Nonlinear  Oscillations, Cambridge University Press, Cambridge, 1981.
%
\bibitem{axel}
A. Pelster, H. Kleinert, and M. Schanz, Phys. Rev. E \textbf{67}, 016604  (2003).
%
\bibitem{axel2}
I. Vidanovic, A. Bala\v{z}, H. Al-Jibbouri, and A. Pelster, Phys. Rev. A \textbf{84}, 013618  (2011).
%
\bibitem{axel3}
H. Al-Jibbouri, I. Vidanovic, A. Bala\v{z}, and A. Pelster, J. Phys. B \textbf{46}, 065303 (2013).

\end{thebibliography}
\end{document}